\documentclass[a4paper,notitlepage,10pt]{article}

\usepackage[ansinew]{inputenc}
\usepackage[lmargin=3cm, rmargin=3cm,bottom=3.15cm,top=3cm]{geometry}
\usepackage{amsfonts}
\usepackage{amsmath}
\usepackage{amsthm}
\usepackage{color}
\usepackage{amssymb}
\usepackage{url}
\usepackage{bm} 
\usepackage{dsfont} 

\usepackage{mathrsfs} 

\usepackage{mathtools}
\mathtoolsset{showonlyrefs}

\usepackage{fixltx2e,amsmath}
\MakeRobust{\eqref}

\newcommand{\opnorm}[1]{\|{#1}\|_\mathrm{op}}
\newcommand{\ER}{{Erd\H{o}s-R\'enyi }}

\newcommand{\Gn}{G_{(n)}}
\newcommand{\In}{\mathbf{I}_{(n)}}
\newcommand{\An}{\mathbf{A}_{(n)}}

\newcommand{\Rn}{\mathbf{R}_{(n)}}
\newcommand{\Rndag}{\mathbf{R}_{(n)}^\dagger }
\newcommand{\Rninv}{\mathbf{R}_{(n)}^{-1} }
\newcommand{\Ln}{\mathbf{L}_{(n)}}
\newcommand{\Yn}{\mathbf{Y}_{(n)}}
\newcommand{\Lnzero}{\mathbf{L}_{(n),0}}
\newcommand{\Lnone}{\mathbf{L}_{(n),1}}
\newcommand{\Dnone}{\mathbf{D}_{(n),1}}
\newcommand{\Anone}{\mathbf{A}_{(n),1}}
\newcommand{\Dn}{\mathbf{D}_{(n)}}
\newcommand{\Lndag}{\mathbf{L}_{(n)}^\dagger}

\newcommand{\bfA}{\mathbf{A}}
\newcommand{\bfD}{\mathbf{D}}
\newcommand{\bfL}{\mathbf{L}}

\newcommand{\Kf}{\mathrm{Kf}}

\newcommand{\expect}{\mathbb{E}}
\newcommand{\expectt}[1]{\mathbb{E}\!\left\{{#1}\right\}}
\newcommand{\trace}{\mathrm{trace}}

\renewcommand{\Pr}{\operatorname{Prob}}
\newcommand{\E}{\expect}

\newcommand{\calY}{\mathcal{Y}}

\newcommand{\transpose}{^\top\! }

\newcommand{\Rnn}{{\mathbb{R}^{n\times n}}}

\newcommand{\diag}{\mathrm{diag}}

\newcommand{\frobnorm}[1]{\left\|{#1}\right\|_\mathrm{F}}

\newcommand{\sqfrobnorm}[1]{\frobnorm{#1}^2}

\definecolor{darkgreen}{rgb}{0 , .5, 0}
\definecolor{lightseagreen}{rgb}{0.1255, 0.6980, 0.6667}

\newcommand{\dist}{\mathrm{dist}}

\newtheorem{theorem}{Theorem}
\newtheorem{lemma}{Lemma}

\newtheorem{assumption}{Assumption}
\newtheorem{example}{Example}

\begin{document}

\title{Concentration of the Kirchhoff index for \ER graphs}

\author{Nicolas Boumal\footnote{Department of mathematical engineering, ICTEAM Institute, Universit\'e catholique de Louvain, Belgium.} \and Xiuyuan Cheng\footnote{Program in Applied and Computational Mathematics, Princeton University, New Jersey, USA.}}

\date{Compiled on: \today.}

\maketitle

\begin{abstract}

Given an undirected graph, the resistance distance between two nodes is the resistance one would measure between these two nodes in an electrical network if edges were resistors. Summing these distances over all pairs of nodes yields the so-called Kirchhoff index of the graph, which measures its overall connectivity.
In this work, we consider \ER random graphs. Since the graphs are random,
their Kirchhoff indices are random variables. We give formulas for the expected value of the Kirchhoff index and show it concentrates around its expectation.
We achieve this by studying the trace of the pseudoinverse of the Laplacian of \ER graphs.
For synchronization (a class of estimation problems on graphs) our results imply that acquiring pairwise measurements uniformly at random is a good strategy, even if only a vanishing proportion of the measurements can be acquired.
\vspace{3mm}
\newline
Keywords: Resistance distance, Kirchhoff index, Erd\H{o}s-R\'enyi, estimation on graphs, synchronization, Cram\'er-Rao bounds, pseudoinverse of graph Laplacian, random matrices, sensor network localization.
\end{abstract}

\section{Introduction}

Consider an undirected, \emph{connected}, weighted graph $G$ with nodes 1 to $n$ and adjacency matrix $\bfA$, such that $A_{ij} = A_{ji} \geq 0$ denotes the weight of the edge connecting nodes $i$ and $j$ (zero if there is no such edge). The degree matrix $\bfD$ is diagonal and such that $D_{ii} = \sum_j A_{ij}$ is the sum of the weights of the edges adjacent to node $i$. A popular notion of distance between two nodes $i$ and $j$ in the graph is the so-called \emph{resistance distance}~\cite{klein1993resistance}:
\begin{align}
	\dist_G(i, j) & = (\bfL^\dagger)_{ii} + (\bfL^\dagger)_{jj} - 2(\bfL^\dagger)_{ij},
\end{align}
where $\bfL$ is the (combinatorial) Laplacian of $G$, defined by $\bfL = \bfD - \bfA$ and $\bfL^\dagger$ denotes its Moore-Penrose pseudoinverse.\footnote{For a symmetric matrix $M$ with eigenvalue decomposition $M = UDU\transpose$, $U\transpose U = I$ and $D = \diag(\lambda_1, \ldots, \lambda_n)$, this pseudoinverse of $M$ is $M^\dagger = UD^\dagger U\transpose$, with $D^\dagger = \diag(\lambda_1^\dagger, \ldots, \lambda_n^\dagger)$. The pseudoinverse of a scalar $\lambda_i$ is $\lambda_i^\dagger = 1/\lambda_i$ if $\lambda_i \neq 0$ and $\lambda_i^\dagger = 0$ if $\lambda_i = 0$.} In an electrical network with $n$ nodes and a resistor of value $1/A_{k\ell}$ across any two nodes $k$ and $\ell$ if they are linked by an edge in $G$, this distance corresponds to the effective electrical resistance one would measure between nodes $i$ and $j$. Interestingly, it is proportional to the average time it takes a random walker to commute between $i$ and $j$~\cite{chandra1996electrical}. The smaller the distance, the better nodes $i$ and $j$ are connected. Klein and Randi{\'c} define the \emph{Kirchhoff index} of the graph $G$ as the sum of all resistance distances~\cite[Thm.\,F]{klein1993resistance}:
\begin{align}
	\Kf(G) & = \sum_{i < j} \dist_G(i, j) = n \cdot \trace(\bfL^\dagger).
\end{align}
A small value indicates a well-connected graph. See the note by Zhou and Trinajsti{\'c}~\cite{zhou2008kirchhoff} for properties of $\Kf(G)$ and its many uses in mathematical chemistry. It is well-known that the spectrum of $\bfL$ captures the connectivity properties of the graph, and it is hence not surprising to see it appear as above. Expander graphs for example, which are both sparse and well-connected, have eigenvalues bounded away from zero (except for one)~\cite{hoory2006expander}. In turn, this translates in small eigenvalues for $\bfL^\dagger$ and a small value of $\Kf(G)$.

For a random graph, the Kirchhoff index is a random variable. In this work, we investigate the random variable $\Kf(G)$ for the case where $G$ is a (connected) \ER random graph, that is, a graph for which each edge has a fixed probability of being present, independently from all others. $\Kf(G)$ can be studied through the spectrum of the random matrix $\bfL$, for which a lot is already known~\cite{ding2010spectral,chung2004spectra,chung2006complex}. As we show below, the Kirchhoff index of large (connected) \ER graphs rapidly concentrates around its expected value, for which we provide formulas. 

A principal motivation for the present investigation is the study of the \emph{synchronization} problem. Synchronization is the task of estimating $n$ elements $g_1, \ldots, g_n$ in a group $\mathcal{G}$ based on certain (not all) relative measurements $h_{ij}\in\mathcal{G}$ which bear information about the ratios $g_i^{} \cdot g_j^{-1}$. The set of measurements defines an undirected graph $G$ on $n$ nodes, with an edge between nodes $i$ and $j$ if a measurement $h_{ij}$ is available. This is an important class of estimation problems on graphs, occurring frequently in applications~\cite{singer2010angular,cucuringu2011eigenvector,wang2012LUD,SondaySingerKevrekidis2013,barooah2007estimation,howard2010estimation,tron2009distributed,cucuringu2011sensor,huang2013consistent,boumal2013MLE,carmona2012analytical,hartley2013rotation}.

Synchronization proves useful both for discrete groups---such as $\mathbb{Z}_2 = \{+1, -1\}$~\cite{cucuringu2011sensor} and the group of permutations~\cite{huang2013consistent}---and for Lie groups---such as the group of translations~\cite{barooah2007estimation,cucuringu2011sensor} and the group of rotations~\cite{boumal2013MLE,carmona2012analytical}.

For the latter two groups,
Cram\'er-Rao bounds (CRB's) were established that put a lower-bound on the variance of any unbiased estimator for these estimation problems~\cite{barooah2007estimation,crbsubquot,crbsynch,howard2010estimation}.
In the case of isotropic, i.i.d.\ noise on the measurements $h_{ij}$, these bounds are proportional to
$\trace(\bfL^\dagger)$, and hence to the Kirchhoff index of $G$ (with weights dictated by the noise distribution), hence the link with our present work. 

As an example, consider synchronization of translations: the group is $\mathbb{R}^d$ and the group operation $\cdot$ is the sum. Let $x_1, \ldots, x_n \in \mathbb{R}^d$ be the $n$ state vectors to estimate. In a sensor network localization context, they could represent the positions of $n$ agents in some coordinate system. For each edge $(i,j)$ in the (fixed, known) graph $G$, $h_{ij}$ is a noisy measurement of the relative position $x_i - x_j$.
Assume they are given by $h_{ij} = -h_{ji} = x_i - x_j + n_{ij}$, with noise $n_{ij} \sim \mathcal{N}(0, \Sigma)$ i.i.d.\ normal random variables. Let $\hat x_1, \ldots, \hat x_n$ be any unbiased estimator of the state vectors. Assuming the $x_i$'s and the $\hat x_i$'s are centered (since the estimation can only be resolved up to a global translation), the CRB for this synchronization problem lower-bounds the variance as~\cite{crbsubquot}:
\begin{align*}
	\expect\Big\{\sum_{i=1}^{n} \|x_i - \hat x_i\|^2\Big\} \geq \trace(\Sigma) \, \trace(\bfL^\dagger).
\end{align*}
The expectation is taken w.r.t.\ the noise $n_{ij}$. The maximum likelihood estimator achieves this bound~\cite{barooah2007estimation}.

Thus, studying the Kirchhoff index of \ER graphs will elucidate the behavior of the CRB on such synchronization tasks where measurements are acquired uniformly at random. For a growing number of nodes $n$, it is desirable to drive the lower-bound on the average variance per node, $\trace(\bfL^\dagger)/n$, to zero (if possible) so as to allow accurate estimation. We will see that this can be achieved even as the edge density decays to zero, provided the graph remains sufficiently connected.

\section{Contribution}

Let $\Gn$ represent an \ER random graph with $n$ nodes and edge presence probability $p_n$. More precisely, for $n \geq 2$ and let $\{A_{ij}^{(n)}\}$, $1 \leq i < j \leq n$, be a collection of independent Bernoulli random variables with
\begin{align}
	A_{ij}^{(n)} & = \begin{cases} 1 & \textrm{ with probability } p_n, \\ 0 & \textrm{ with probability } 1-p_n, \end{cases}
	\label{eq:ER}
\end{align}
for some edge probability $0 < p_n < 1$ which may depend on $n$. Because we consider simple, undirected graphs, define $A_{ji}^{(n)} = A_{ij}^{(n)}$ and $A_{ii}^{(n)} = 0$.
For $i \neq j$,
\begin{align}
	& \expect\{A_{ij}^{(n)}\} = p_n, \\ & \expect\{(A_{ij}^{(n)}-p_n)^2\} = p_n(1-p_n) \triangleq \sigma_n^2 .
\end{align}
We let $\An$ be the $n\times n$ adjacency matrix of $\Gn$:
\begin{align}
	\An & =
\begin{pmatrix}
	0      & A_{12}^{(n)} & \cdots & A_{1n}^{(n)} \\
	A_{21}^{(n)} & 0      & \cdots & A_{2n}^{(n)} \\
	\vdots & \vdots & \ddots & \vdots \\
	A_{n1}^{(n)} & A_{n2}^{(n)} & \cdots & 0
\end{pmatrix}.
\end{align}
Let $\mathds{1}_n$ denote the all-ones vector of length $n$. The diagonal degree matrix is given by $\Dn = \diag(\An \mathds{1}_n)$. Then, the $n\times n$ Laplacian matrix is defined as $\Ln = \Dn - \An$:
\begin{align}
	\Ln & = 
\begin{pmatrix}
	\sum_{i\neq 1} A_{1i}^{(n)} & -A_{12}^{(n)} & \cdots & -A_{1n}^{(n)} \\
	-A_{21}^{(n)} & \sum_{i\neq 2} A_{2i}^{(n)}      & \cdots & -A_{2n}^{(n)} \\
	\vdots & \vdots & \ddots & \vdots \\
	-A_{n1}^{(n)} & -A_{n2}^{(n)} & \cdots & \sum_{i\neq n} A_{ni}^{(n)}
\end{pmatrix}.
\end{align}
The Laplacian is a symmetric, positive semidefinite matrix since for all $u\in\mathbb{R}^n$,
\begin{align}
	u\transpose \Ln u = \sum_{i<j} A_{ij}^{(n)} (u_i - u_j)^2 \geq 0.
\end{align}
Its eigenvalues $\lambda_i^{(n)}$, $1 \leq i \leq n$, are thus real and nonnegative. Furthermore, since $\Ln\mathds{1}_n = 0$, the smallest eigenvalue is necessarily equal to zero:
\begin{align}
	0 = \lambda_1^{(n)} \leq \lambda_2^{(n)} \leq \cdots \leq \lambda_n^{(n)}.
\end{align}

Notice that for all $n$ and $p_n < 1$, there is a strictly positive probability that $\Gn$ is disconnected. By definition,
\begin{align}
	\Kf(\Gn) & = \begin{cases}
	n \cdot \trace(\Lndag) & \textrm{if } \Gn \textrm{ is connected,} \\
	\infty & \textrm{otherwise.}
	\end{cases}
\end{align}
Then, for all $n$, the expectation $\expectt{\Kf(\Gn)}$ is necessarily infinite, which is not very interesting. Instead, we study the expectation of $\Kf(\Gn)$ restricted to \emph{connected} \ER graphs. Let $C_n$ denote the event that $\Gn$ is connected, let $\lnot C_n$ denote the event that $\Gn$ is disconnected and let $I_E$ denote the indicator function for an event $E$, that is, $I_E$ evaluates to 1 if the event $E$ is realized, 0 otherwise. Then, our quantity of interest is:
\begin{align}
	\expectt{\Kf(\Gn) I_{C_n}} = \expectt{n \cdot \trace(\Lndag) I_{C_n}} = n \cdot \expectt{\trace(\Lndag)} - n \cdot \expectt{\trace(\Lndag) I_{\lnot C_n}}.
\end{align}
In particular,
\begin{align}
	\left| \expectt{\Kf(\Gn) I_{C_n}} - n \cdot \expectt{\trace(\Lndag)} \right| \leq n \cdot \max_{G} \trace(\bfL^\dagger) \cdot \Pr\left[ \lnot C_n \right],
	\label{eq:legit}
\end{align}
where $\Pr[E]$ is the probability of event $E$. The maximum is taken over all graphs $G$ with $n$ nodes, and $\bfL$ is the Laplacian of $G$. This maximum is quadratic in $n$.
\begin{lemma}
	$\max_{G} \trace(\bfL^\dagger) = \frac{n^2-1}{6}$.
	\label{lem:maxtrace}
\end{lemma}
Thus, if we make $\Pr\left[ \lnot C_n \right]$ go to zero sufficiently fast, $n \cdot \expectt{\trace(\Lndag)}$ is a good proxy for $\expectt{\Kf(\Gn) I_{C_n}}$. It is well known that if $np_n \gg \log n$, then $G$ is asymptotically almost surely connected. The notation $f(n) \gg g(n)$ is used to mean $f(n)/g(n) \to \infty$ as $n\to\infty$. We ask for slightly more than sheer connectivity:
\begin{assumption}
$np_n \gg \log^6 n$.
\label{assu:large}
\end{assumption}
Indeed, under Assumption~\ref{assu:large}, not only is the graph $\Gn$ asymptotically almost surely connected, but also the $n-1$ nonzero eigenvalues of $\Ln$ all concentrate around the same value, far from zero. We come back to this momentarily.

The (appropriately scaled) scalar random variable
\begin{align}
	X_n & = p_n \trace(\Lndag)
	\label{eq:Xn}
\end{align}
is the variable of interest from now on. We may gain a quick insight into this variable by considering the expected value of $\Ln$:
\begin{align}
	\Lnzero \triangleq \expectt{\Ln} & = p_n(n\In - \mathds{1}_{n\times n}),
\end{align}
where $\In$ denotes the $n\times n$ identity matrix and $\mathds{1}_{n\times n}$ is the all-ones matrix of size $n\times n$. The eigenvalues of $\Lnzero$ are $0, n p_n, n p_n, \ldots, n p_n$. 
Thus, $p_n\trace(\Lnzero^\dagger) = (n-1)/n$, which goes to $1$ as $n\to\infty$. Indeed, we will see that as $n$ goes to infinity, $\Ln$ ``behaves more and more like $\Lnzero$,'' so that $X_n \to 1$. We aim at a more precise statement that will already be useful for moderate $n$.

In the limit, the positive eigenvalues of $\Ln$ tend to be distributed symmetrically around $np_n$~\cite{ding2010spectral}. Thus, the positive eigenvalues of $\Lndag$ will be distributed asymmetrically around $1/np_n$ and the expected value of $X_n$ will be biased away from 1. In what follows, we establish a formula for the expected value of $X_n$ for large $n$ and bound its fluctuation around its mean.

To this end, we look at the spectrum of the centered Laplacian. Define these centered random variables:
\begin{align}
	X_{ij}^{(n)} & \triangleq  A_{ij}^{(n)}-p_n .
\end{align}
Then, $\expect\{X_{ij}^{(n)}\} = 0$ and $\expect\{(X_{ij}^{(n)})^2\} = \sigma_n^2$. Define the $n\times n$ matrix $\Lnone$ such that $\Ln =  \Lnzero + \Lnone$:
\begin{align}
	\Lnone & =
\begin{pmatrix}
	\sum_{i\neq 1} X_{1i}^{(n)} & -X_{12}^{(n)} & \cdots & -X_{1n}^{(n)} \\
	-X_{21}^{(n)} & \sum_{i\neq 2} X_{2i}^{(n)}      & \cdots & -X_{2n}^{(n)} \\
	\vdots & \vdots & \ddots & \vdots \\
	-X_{n1}^{(n)} & -X_{n2}^{(n)} & \cdots & \sum_{i\neq n} X_{ni}^{(n)}
\end{pmatrix}.
\end{align}
Assumption~\ref{assu:large} ensures the eigenvalues of $\Lnone/np_n$ concentrate around 0, which ensures connectivity and limits the fluctuation of $X_n$ around its mean. Indeed, for a symmetric matrix $M \in \Rnn$, define the operator norm $\opnorm{M}$ of $M$ as
\begin{align}
	\opnorm{M} & \triangleq \max_{u\in\mathbb{R}^n, u\transpose u = 1} \sqrt{u\transpose M\transpose M u} = \max_{1 \leq i \leq n} |\lambda_i(M)|,
\end{align}
and define the event
\begin{align}
	E_n & \textrm{ is the event that } \opnorm{\Lnone} \leq 5\sqrt{np_n \log n}.
	\label{eq:En}
\end{align}
This event happens with high probability.
\begin{lemma}
	Under Assumption~\ref{assu:large}, there exists $n_0$ such that, for all $n\geq n_0$,
	\begin{align}
		\Pr[E_n] & \geq 1 - \frac{3.01}{n^{11}}.
	\end{align}
	\label{lem:PrEn}
\end{lemma}
The proof of this lemma rests essentially upon results from Chung et al.~\cite{chung2004spectra}. Notice that if $\Gn$ is not connected, then $\opnorm{\Lnone} \geq np_n$ and in particular $E_n$ is false. Hence,
\begin{align}
	\Pr[\lnot C_n] \leq \Pr[\lnot E_n] \leq \frac{3.01}{n^{11}}.
\end{align}
Then, lemmas~\ref{lem:maxtrace} and~\ref{lem:PrEn} combined indeed show that, under Assumption~\ref{assu:large}, the right hand side of~\eqref{eq:legit} decays at least as fast as $1/n^8$; thus justifying the study of $\trace(\Lndag)$.

Under Assumption~\ref{assu:large}, we establish formulas for the expectation of $X_n$:
\begin{theorem}[Expectation] \label{thm:expectation}
	Under Assumption~\ref{assu:large},
	\begin{align}
			\expect{X_n} & = 1 + \left(2\frac{1-p_n}{p_n} - 1\right)\frac{1}{n} + \mathcal{O}\left( \frac{\log^2 n}{(np_n)^2} \right),
		\label{eq:generalbound}
	\end{align}
	with $X_n = p_n \trace(\Lndag)$, see eq.~\eqref{eq:Xn}.
\end{theorem}
Furthermore, $X_n$ concentrates around its mean:
\begin{theorem}[Fluctuation]\label{thm:X-EX}
Under Assumption~\ref{assu:large}, for all $0 < \epsilon \leq 1/2$, there exists $n_0$ such that for all $n \geq n_0$, it holds with probability at least $1- 2\epsilon - 3.01/n^{4}$ that:
\begin{align}
	| X_n - \E X_n | & \leq 2.02 \frac{\sqrt{\log (1/\epsilon) }}{np_n}.
\end{align}
The constant 2.02 can be made arbitrarily close to 2 for large enough $n$.
\end{theorem}

We spell out two simple examples. In both cases, Assumption~\ref{assu:large} is readily checked.

\begin{example}[$p_n$ bounded away from 0 and 1]\label{examplebounded}
Let the edge presence probability $p_n$ remain bounded away from zero and one, i.e., there exist constants $c_\ell, c_u$ such that $0 < c_\ell \leq p_n \leq c_u < 1$ for all $n$. Then,
\begin{align}
	\expect{X_n} = 1 + \left(2\frac{1-p_n}{p_n} - 1\right)\frac{1}{n} + \mathcal{O}\left( \frac{\log^2{n}}{n^2} \right).
	\label{eq:ERboundedpequation}
\end{align}
Furthermore, for all $0 < \epsilon \leq 1/2$, for $n$ large enough, with probability at least $1-2\epsilon - 3.01/n^{4}$,
\begin{align}
	|X_n-\expect{X_n}| & \leq 2.02 \frac{\sqrt{\log(1/\epsilon)}}{np_n} = \mathcal{O}\left(\frac{1}{n}\right).
\end{align}
\end{example}

\begin{example}[vanishing $p_n$]\label{examplevanishing}
Let the edge presence probability $p_n$ decay as $p_n = \gamma n^{-1+\alpha}$, for some constant $\gamma > 0$ and $0 < \alpha \leq 1$ (and $\gamma < 1$ if $\alpha = 1$). Then,
\begin{align}
	\expect{X_n} & = 1 + \frac{2}{\gamma n^\alpha} - \frac{3}{n} + \mathcal{O}\left( \frac{\log^2{n}}{n^{2\alpha}} \right).
	\label{eq:example2mean}
\end{align}
Furthermore, for all $0 < \epsilon \leq 1/2$, for $n$ large enough, with probability at least $1-2\epsilon - 3.01/n^{4}$,
\begin{align}
	|X_n-\expect{X_n}| & \leq 2.02 \frac{\sqrt{\log(1/\epsilon)}}{\gamma n^\alpha} = \mathcal{O}\left(\frac{1}{n^\alpha}\right).
	\label{eq:example2fluctuation}
\end{align}
\end{example}

\begin{figure*}[t]
\centering
\includegraphics[width=1\linewidth]{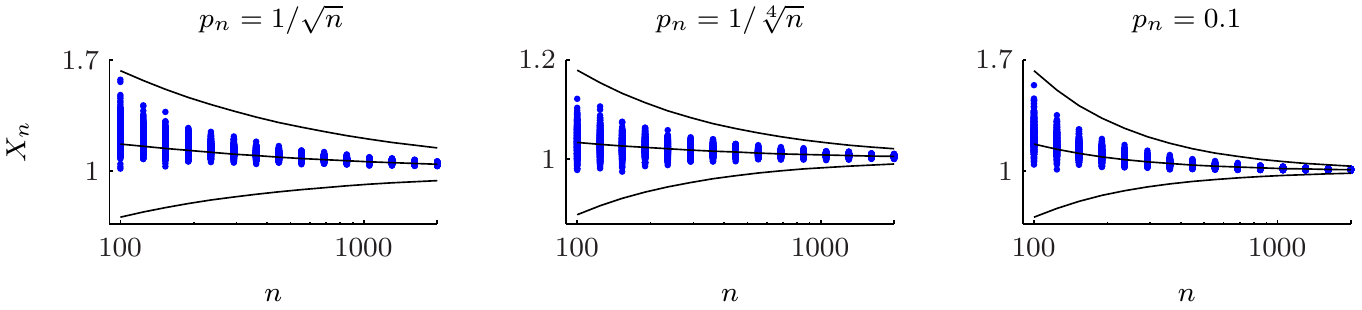}
\caption{For a number of nodes $n$ ranging from 100 to 2000 on a log-scale (15 different values) and for three different scenarios of edge densities $p_n$ (decaying as $1/\sqrt{n}$, decaying as $1/\sqrt[4]{n}$ and constant), we computed 500 realizations of $X_n = p_n\trace(\Lndag)$ (blue dots, $3\times 15\times 500$ in total). The middle black curve is the established formula for $\expect{X_n}$~\eqref{eq:example2mean}, ignoring higher order terms. The black curves above and below are separated from the middle one by the bound on the fluctuation~\eqref{eq:example2fluctuation}, such that, for large $n$, all blue dots are within the two black lines with probability at least 99\%. These bounds appear to make sense already for reasonable values of $n$. 
}
\label{fig:tracepinvlap1}
\end{figure*}

All the aforementioned results have a ``for large $n$'' proviso, which in theory limits their usability. Through numerical tests, Figure~\ref{fig:tracepinvlap1} illustrates the fact that already for small values of $n$ (in the hundreds), the predictions hold quite well.

With respect to the Kirchhoff index, these results show that, under Assumption~\ref{assu:large}, the Kirchhoff index of a connected \ER graph on $n$ nodes, with $n$ large enough, concentrates around its expectation, which is given by:
\begin{align}
		\expectt{\Kf(\Gn) I_{C_n}} & = \frac{1}{p_n}n + \left[ \frac{2}{p_n^2} - \frac{3}{p_n}\right]  +  \mathcal{O}\left(\frac{1}{p_n^3}\frac{\log^2 n}{n}\right).
\end{align}
The difference to the mean is, with high probability, on the order of $\mathcal{O}(1/p_n^2)$.

With respect to synchronization, the take home message of Example~\ref{examplevanishing} is the following. If we could afford a complete graph of measurements, the Cram\'er-Rao lower-bound on the average variance per node would decay as $\trace(\Lndag)/n = \mathcal{O}(1/n)$. If on the other hand we can only afford to obtain a vanishing fraction of, say, $p_n \sim 1/\sqrt{n}$ of the $\mathcal{O}(n^2)$ measurements following an \ER graph---which is easy to set up in a decentralized manner---then we would have that $\expect\big\{\trace(\Lndag)\big\}/n = \expect X_n / np_n = \mathcal{O}(1/\sqrt{n})$ and that this quantity concentrates around its mean as $\mathcal{O}(1/n)$. Hence, although we only acquire a vanishing fraction of the measurements, acquiring them uniformly at random guarantees sufficient connectivity in the graph that the average CRB still goes to zero as $n\to\infty$, with a fairly predictable price to pay in accuracy.

The remainder of the paper presents proofs for lemmas~\ref{lem:maxtrace} and~\ref{lem:PrEn} and for theorems~\ref{thm:expectation} and~\ref{thm:X-EX}.

\paragraph{Further notation} We denote the (random) eigenvalues of $\Lnone$ as
\begin{align}
	t_1^{(n)} = 0, \quad t_2^{(n)} \leq \cdots \leq t_n^{(n)}.
\end{align}
These can be either positive or negative. The eigenvalues of $\Ln$ obey
\begin{align}
	\lambda_1^{(n)} & = 0, \quad \lambda_{i>1}^{(n)} = n p_n +  t_i^{(n)}.
\end{align}
For a matrix $M$, $\frobnorm{M} = \sqrt{\trace(M\transpose M)}$ denotes its Frobenius norm.

\section{Proof of Lemma~\ref{lem:maxtrace} about $\max \trace(\bfL^\dagger)$}

We here argue that for any simple, undirected, unweighted graph $G$ with $n$ nodes (that is, $A_{ij} = A_{ji} \in \{0, 1\}$ and $A_{ii} = 0$), letting $\bfL$ be its Laplacian matrix, it holds that $\trace(\bfL^\dagger) \leq (n^2-1)/6$. Furthermore, this inequality is tight, as it is attained if $G$ is a chain (i.e., a path).

Let us first consider the case of $G$ connected. Klein and Randi{\'c}~\cite[Thm.\,D]{klein1993resistance} state that the resistance distance between any two nodes $i$ and $j$, $\dist_G(i, j)$, is bounded by the shortest path distance between these nodes, $\dist_{G,\textrm{sp}}(i, j)$. Thus,
\begin{align}
	\Kf(G) = \sum_{i < j} \dist_G(i, j) \leq \sum_{i < j} \dist_{G,\textrm{sp}}(i, j) \triangleq \operatorname{W}(G), 
\end{align}
where $\operatorname{W}(G)$ is the so-called \emph{Wiener index} of $G$. Entringer et al.~\cite[Thm.\,2.3]{entringer1976distance} show that $\operatorname{W}(G) \leq n(n^2-1)/6$ and that this value is attained if $G$ is a path graph. (An easier bound is found by noting that $\dist_{G,\textrm{sp}}(i, j) \leq n-1$ and there are on the order of $n^2$ terms in the sum.) Since $\trace(\bfL^\dagger) = \Kf(G)/n$, this proves the lemma for connected graphs.

Now let $G$ have $k$ connected components, and let $G_1, \ldots, G_k$ be the graphs corresponding to these components, such that $G_i$ has $n_i$ nodes and $n_1 + \cdots + n_k = n$. Furthermore, let $\bfL_1, \ldots, \bfL_k$ be the Laplacian matrices associated to these components. Without loss of generality, the $n$ nodes may be renumbered (permuted) such that $\bfL$ is block diagonal, with blocks given by the $\bfL_i$'s. Thus,
\begin{align}
	\trace(\bfL^\dagger) = \sum_{i=1}^k \trace(\bfL_i^\dagger) \leq \sum_{i=1}^k \frac{n_i^2 - 1}{6} \leq \frac{n^2 - 1}{6},
\end{align}
which concludes the proof.

\section{Proof of Lemma~\ref{lem:PrEn} about the probability of $E_n$}

Throughout, we let Assumption~\ref{assu:large} hold. We prove that for any $\epsilon_0 > 0$ and for any $\alpha \geq 1$, there exists $n_0$ such that for all $n \geq n_0$,
\begin{align}
	\Pr\left[ \opnorm{\Lnone} \geq \sqrt{2(\alpha+1+\epsilon_0)} \sqrt{np_n\log{n}} \right] \leq \frac{3.01}{n^\alpha}.
	\label{eq:specnormMvanishing}
\end{align}
Lemma~\ref{lem:PrEn} follows from setting $\alpha = 11$ and $\epsilon_0 = 1/2$.

Recall that $\Lnone = \Ln - \expectt{\Ln}$. Define similarly
\begin{align}
	\Dnone & = \Dn - \expectt{\Dn} = \Dn - (n-1)p_n \In, \\ \Anone & = \An - \expectt{\An} = \An - p_n (\mathds{1}_{n\times n} - \In),
\end{align}
so that $\Lnone = \Dnone - \Anone$.

We first show that for all $\epsilon_1 > 0$ and for all $\alpha \geq 1$, there exists $n_0$ such that for all $n \geq n_0$,
\begin{align}
	\Pr\left[ \opnorm{\Anone} \geq (2+\epsilon_1) \sqrt{np_n} \right] \leq \frac{1.01}{n^\alpha}.
	 \label{eq:Apart}
\end{align}
Then we show that for all $\epsilon_2 > 0$ and for all $\alpha \geq 1$, there exists $n_0$ such that for all $n \geq n_0$,
\begin{align}
	\!\!\!\!\!\!\!\!\Pr\left[  \opnorm{\Dnone} \geq \sqrt{2(\alpha+1+\epsilon_2)} \sqrt{np_n\log{n}} \right] \leq \frac{2}{n^\alpha}.
	\label{eq:Dpart}
\end{align}
Then, since $\opnorm{\Lnone} \leq \opnorm{\Anone} + \opnorm{\Dnone}$, the proof will be complete (union bound).

We first consider $\Anone$ and resort to~\cite[Thm.~3.2]{chung2004spectra}. In the latter paper, self loops in the graph are allowed. To take this into account, let $Y_i^{(n)}$, $n\geq 2$, $1 \leq i \leq n$, be i.i.d.\ Bernoulli random variables with the same distribution as, and independent from, the $A_{ij}^{(n)}$'s, $i < j$, and let $\Yn = \diag(Y_1^{(n)}, \ldots, Y_n^{(n)})$. Then, in the present work's notation, reference~\cite{chung2004spectra} defines a matrix $C$ such that
\begin{align}
	np_n C = \Anone + (\Yn - p_n\In).
	\label{eq:Aneq1}
\end{align}
Since $np_n \gg \log^6 n$, there exists a function $g$ such that $g(n) \to \infty$ and such that $np_n \gg (g(n)\log n)^6$. Inspecting the proof of Theorem~3.2 in that reference, we see that it is shown that, for all $\epsilon_1 > 0$, for large enough $n$,
\begin{align}
	\Pr\big[ \opnorm{np_n C} & \geq (2+\epsilon_1)\sqrt{np_n} \big] \\
		& \leq \frac{1.01 \cdot n}{(1+\epsilon_1)^{2g(n)\log n}} \\
		& = 1.01 \cdot n^{1-2\log(1+\epsilon_1)g(n)}.
	\label{eq:Aneq2}
\end{align}
Since $g(n) \to \infty$, this decays faster than any polynomial $\mathcal{O}(1/n^\alpha)$. The contribution of $\Yn-p_n\In$ is negligible:
\begin{align}
	\forall n\geq 2, \ \opnorm{\Yn-p_n\In} & = \max_{1 \leq i \leq n} |Y_i^{(n)}-p_n| \leq 1.
	\label{eq:Aneq3}
\end{align}
Combining the three last equations shows~\eqref{eq:Apart}.

Now consider $\opnorm{\Dnone} = \max_{1\leq i \leq n} |(\Dnone)_{ii}|$. Each diagonal entry
\begin{align}
	(\Dnone)_{ii} & = \sum_{j\neq i} X_{ij}^{(n)} = \Big( \sum_{j\neq i} A_{ij}^{(n)} \Big) - (n-1)p_n
\end{align}
is the difference between a sum of independent, nonnegative Bernoulli random variables and its mean. The Chernoff bound for nonnegative Bernoulli's~\cite[Thm.~2.4]{chung2006complex} controls such differences as follows, for positive $\lambda$:
\begin{align}
\Pr\left[ (\Dnone)_{ii} \geq +\lambda \right] & \leq \exp\left( -\frac{\lambda^2 }{2((n-1)p_n + \frac{\lambda}{3})} \right),  \\
\Pr\left[ (\Dnone)_{ii} \leq -\lambda \right] & \leq \exp\left( -\frac{\lambda^2 }{2(n-1)p_n} \right).
\end{align}
Combining these two inequalities and setting $\lambda = \sqrt{ c np_n \log{n} }$ for some positive $c$ yet to determine, we obtain:
\begin{align}
& \Pr \left[ \big|(\Dnone)_{ii}\big| \geq \sqrt{ c np_n \log{n} }\right] \\
	& \quad \leq \exp\left( -\frac{c np_n \log{n}}{2\left((n-1)p_n + \frac{\sqrt{ c np_n \log{n} }}{3}\right)} \right) + \\ & \quad\quad\, \exp\left( -\frac{c np_n \log{n}}{2(n-1)p_n} \right) \nonumber\\
	& \quad \leq \exp\left( -\frac{c}{2} \frac{\log{n}}{\frac{n-1}{n} + \frac{1}{3}\sqrt{\frac{ c \log{n} }{np_n} }} \right) + \exp\left( -\frac{c}{2} \log{n} \right).
\end{align}
Given that $\sqrt{c \log{n} / np_n} \to 0$ as $n\to\infty$, for all $\epsilon_2 > 0$, there exists $n_{\epsilon_2,c}$ such that for all $n \geq n_{\epsilon_2,c}$, the denominator in the first exponential obeys $\frac{n-1}{n} + \frac{1}{3}\sqrt{\frac{ c \log{n} }{np_n} } \leq 1+\epsilon_2$. Hence, we further obtain:
\begin{align}
& \Pr \left[ \big|(\Dnone)_{ii}\big| \geq \sqrt{ c np_n \log{n} }\right] \\ 
	    & \quad \leq \exp\left( -\frac{c}{2} \frac{\log{n}}{1+\epsilon_2 } \right) + \exp\left( -\frac{c}{2} \log{n} \right) \\
		& \quad \leq 2\exp\left( -\frac{c}{2} \frac{\log{n}}{1+\epsilon_2 } \right) = \frac{2}{n^{\frac{c}{2(1+\epsilon_2)}}}.
\end{align}
By the union bound, which states that the probability of at least one event among $n$ events to occur is bounded by the sum of the probabilities of those $n$ events, it follows that:
\begin{align}
\Pr \left[ \opnorm{\Dnone} \leq \sqrt{ c np_n \log{n} } \right] & \geq 1 - \frac{2}{n^{\frac{c}{2(1+\epsilon_2)}-1}}.
\end{align}
Let $c = 2(1+\epsilon_2)(\alpha+1)$. This proves~\eqref{eq:Dpart} and thus concludes the proof.
 
\section{Proof of Theorem~\ref{thm:expectation} about the expectation}

In this proof of Theorem~\ref{thm:expectation}, we establish an expression for the expectation of $X_n$~\eqref{eq:Xn}. First, remember the definition of event $E_n$~\eqref{eq:En}. Using $I$ to denote the indicator function, we have
\begin{align}
	\expectt{X_n} & = \expectt{X_n I_{E_n}} + \expectt{X_n I_{\lnot E_n}}.
\end{align}
Using lemmas~\ref{lem:maxtrace} and~\ref{lem:PrEn} (under Assumption~\ref{assu:large}), we see that the second term is small. Indeed, for large enough $n$, it holds that
\begin{align}
	0 \leq \expectt{X_n I_{\lnot E_n}} & \leq p_n \frac{n^2-1}{6} \Pr[\lnot E_n] = \mathcal{O}\left(\frac{1}{n^{9}}\right).
\end{align}
Thus, we need only concentrate on the expectation of $X_n$ under the event $E_n$.

Consider the sequence
\begin{align}
	c_n & = 5 \sqrt{\frac{\log n}{np_n}}.
\end{align}
Under event $E_n$, we have $\frac{1}{np_n}\opnorm{\Lnone} \leq c_n$. Furthermore, Assumption~\ref{assu:large} guarantees that $c_n \to 0$ as $n \to \infty$. In particular, for all $n$ larger than some threshold, $c_n < 1$. Recall that the $t_i^{(n)}$'s denote the eigenvalues of $\Lnone$. For $n$ large enough then, $|t_i ^{(n)}| \leq np_nc_n < np_n$, so that $\lambda_2^{(n)} = np_n + t_2^{(n)} > 0$. This means that the graph defined by the adjacency matrix $\An$ is connected.
When such is the case, $X_n$ obeys
\begin{align}
	X_n & = \sum_{i=2}^n \frac{p_n}{\lambda_i^{(n)}} = \sum_{i=2}^n \frac{p_n}{np_n + t_i^{(n)}} = \frac{1}{n} \sum_{i=2}^n \frac{1}{1 + \frac{1}{np_n}\, t_i^{(n)}}.
\end{align}
Then, using the series expansion $\frac{1}{1+x} = 1 - x + x^2 - x^3 + x^4 - \cdots$, which is convergent for $|x| < 1$, we get:
\begin{align}
	X_n & = \frac{1}{n} \sum_{i=2}^n \left[ 1 + \sum_{k=1}^\infty \left(-\frac{1}{np_n}\, t_i^{(n)}\right)^k \right] \\
							& = \frac{1}{n} \left[ (n-1) + \sum_{k=1}^\infty \sum_{i=2}^n \left(-\frac{1}{np_n}\, t_i^{(n)}\right)^k \right].
\end{align}
The summations commute because the series is absolutely convergent. Observe that
\begin{align}
	\sum_{i=2}^n (t_i^{(n)})^k = \trace(\Lnone^k).
\end{align}
Thus, under event $E_n$,
\begin{align}
	X_n & = \frac{1}{n} \left[ (n-1) + \sum_{k=1}^\infty \left(-\frac{1}{np_n}\right)^k \, \trace(\Lnone^k) \right].
	\label{eq:XnIfEn}
\end{align}
In order to compute the expectation of $X_nI_{E_n}$ then, we must understand the expectations $\expect\{\trace(\Lnone^k)I_{E_n}\}$ for $k = 1, 2, \ldots$ As it is easier to compute $\expect\{\trace(\Lnone^k)\}$, we first observe the following. Since $|X_{ij}^{(n)}| \leq 1$, Gershgorin's theorem~\cite[Thm.\,7.2.1]{golub2012matrix} tells us that $\opnorm{\Lnone} \leq 2(n-1)$. Thus, $\trace(\Lnone^k) \leq n \cdot (2(n-1))^k = \mathcal{O}(n^{k+1})$. Using Lemma~\ref{lem:PrEn} again, we find that
\begin{align}
	\expectt{\trace(\Lnone^k)I_{E_n}} & = \expectt{\trace(\Lnone^k)} + \mathcal{O}(n^{k-10}).
\end{align}
Owing to independence, the first few terms are given by:
\begin{align}
	\expectt{\trace(\Lnone^1)} & = \expect\Big\{\sum_{i\neq j} X_{ij}^{(n)}\Big\} = 0, \\
	\expectt{\trace(\Lnone^2)} & = \expectt{\sqfrobnorm{\Lnone}} = 2n(n-1)\sigma_n^2, \\
	\expectt{\trace(\Lnone^3)} & = 0.
\end{align}
So, continuing from~\eqref{eq:XnIfEn}, taking expectations and tracking the error terms:
\begin{align}
	\expectt{X_n I_{E_n}} & = \frac{1}{n} \Big[ (n-1) + 2(n-1)\frac{\sigma_n^2}{np_n^2} + \phantom{x}\\ & \quad\quad \sum_{k=4}^\infty \left(-\frac{1}{np_n}\right)^k \, \expectt{\trace(\Lnone^k) I_{E_n}} \Big] \\ & \quad\quad + \mathcal{O}\left( \frac{1}{n^8 (np_n)^3} \right) \\
	    & = 1 + \left(2\frac{\sigma_n^2}{p_n^2} - 1\right)\frac{1}{n} - 2\frac{\sigma_n^2}{p_n^2}\frac{1}{n^2} + \phantom{x}\\ & \quad\quad \frac{1}{n}\sum_{k=4}^\infty \left(-\frac{1}{np_n}\right)^k \, \expectt{\trace(\Lnone^k) I_{E_n}} \\ & \quad\quad + \mathcal{O}\left( \frac{1}{n^8 (np_n)^3} \right).
\label{eq:expecttrace1}
\end{align}
Let us bound the series:
\begin{align}
	& \left| \frac{1}{n}\sum_{k=4}^\infty \left(-\frac{1}{np_n}\right)^k \, \expectt{\trace(\Lnone^k) I_{E_n}} \right| \\
	& \quad \leq \frac{1}{n} \sum_{k=4}^\infty \frac{1}{(np_n)^k} \, \expectt{\left| \trace(\Lnone^k) I_{E_n} \right| } \\
	& \quad \leq \sum_{k=4}^\infty \frac{1}{(np_n)^k} \, \expectt{ \opnorm{\Lnone}^k I_{E_n} } \\
	& \quad \leq \sum_{k=4}^\infty c_n^k = c_n^4 \sum_{k=0}^\infty c_n^k = \frac{c_n^4}{1-c_n} = \mathcal{O}(c_n^4).
\end{align}
Plugging the latter in~\eqref{eq:expecttrace1} yields~\eqref{eq:generalbound}.

\section{Proof of Theorem~\ref{thm:X-EX} about the fluctuation}

We now show that $X_n$ concentrates around its expected value---Theorem~\ref{thm:X-EX}. The proof rests on an extension of McDiarmid's inequality~\cite{mcdiarmid1998concentration} by Kutin~\cite[Cor.\,3.4]{kutin2002extensions}, which can be stated as follows:
\begin{lemma}[Extended McDiarmid's inequality]\label{lem:mcdiarmid}
Let $\mathbf{Y} = (Y_1, \ldots, Y_m)$ with the $Y_i$'s independent random variables taking values in the set $\calY$. Let $B\subset\calY^m$ be a ``bad'' subset of $\calY^m$. Let $\mathbf{y} = (y_1, \ldots, y_m) \in \calY^m$ and let $\mathbf{y'}\in\calY^m$ such that $\mathbf{y}$ and $\mathbf{y'}$ differ only in one entry. Let $f \colon \calY^m \to \mathbb{R}$ be a measurable function such that for all such pairs $\mathbf{y}, \mathbf{y'}$, it holds that
\begin{align}
	\left| f(\mathbf{y}) - f(\mathbf{y'}) \right| & \leq \begin{cases} c & \textrm{if } \mathbf{y} \notin B, \\ b & \textrm{otherwise.} \end{cases}
\end{align}
Then, for all $\lambda > 0$ and for all $\alpha > 0$:
\begin{align}
	\Pr\big[ \left| f(\mathbf{Y}) - \expect f(\mathbf{Y}) \right| \geq \lambda \big] & \leq 2\exp\left( \frac{-\lambda^2}{2m(c+b\alpha)^2} \right) + \frac{2m}{\alpha}\delta ,
\end{align}
where $\delta \geq \Pr\left[ \mathbf{Y} \in B \right]$ is an upper bound on the probability of bad realizations.
\end{lemma}
In our setting, the $m = n(n-1)/2$ independent random variables $\mathbf{Y}$ are the $X_{ij}$'s. These determine $\Lnone$ compeltely, and we let $f(\Lnone) = X_n$. Modifying just one variable $X_{ij}$ results in a new matrix $\Lnone' = \Lnone \pm E_{ij}$, where $E_{ij} = (e_i - e_j)(e_i - e_j)\transpose$ and $e_i$ denotes the $i^\mathrm{th}$ canonical basis vector of length $n$. In order to determine $c$ in the above lemma, we must bound the difference $|f(\Lnone) - f(\Lnone')|$. In order to obtain a useful bound, we restrict $\Lnone$ to satisfy $E_n$ when determining $c$. With a slight abuse of notation, we write: $\Lnone \in E_n$. Thus, the ``bad'' set $B$ in Lemma~\ref{lem:mcdiarmid} corresponds to $\lnot E_n$. From Lemma~\ref{lem:maxtrace}, we immediately see that, for any realization of $\Lnone$, $|f(\Lnone) - f(\Lnone')| \leq (n^2-1)/6$, so that $b = n^2/6$ is a conservative choice. From Lemma~\ref{lem:PrEn}, we obtain that $\delta = 3.01/n^{11}$ is acceptable for large enough $n$.

We now determine $c$. We first express the function $X_n$ in a different form, where $k$ is the (random) number of connected components of the graph $\Gn$:
\begin{align}
	f(\Lnone) & = X_n = p_n \trace(\Lndag) = p_n \sum_{i=k+1}^{n} \frac{1}{\lambda_i^{(n)}} \\
		    & = p_n \sum_{i=k+1}^{n} \frac{1}{np_n + t_i^{(n)}} \frac{1/np_n}{1/np_n} \\
			& = \frac{1}{n} \sum_{i=k+1}^{n} \frac{1}{1 + \frac{t_i^{(n)}}{np_n}} = \frac{1}{n} \Big( \trace(\Rndag) - 1 \Big),
\end{align}
where $\Rn$ is defined as
\begin{align}
	\Rn & = \In + \frac{1}{np_n} \Lnone.
\end{align}
Indeed, the eigenvalues of $\Rn$ are $1+t_i^{(n)}/np_n, i = 1\ldots n$, with $t_1^{(n)} = 0$. Similarly, we let $\Rn' = \In + \frac{1}{np_n} \Lnone' = \Rn \pm E_{ij}/np_n$.

Notice that, since $\opnorm{E_{ij}} = 2$ and since $\Lnone \in E_n$, by the triangular inequality, it holds that $\opnorm{\Lnone'} \leq \opnorm{\Lnone} + 2 \leq 5.01\sqrt{np_n\log n}$ for large enough $n$. In particular, both $\Rn$ and $\Rn'$ are invertible, so that their pseudoinverse is their inverse. Indeed, the eigenvalues of $\Rn'$ obey
\begin{align}
	|1 - \lambda_i(\Rn')| \leq 5.01 \sqrt{\frac{\log n}{n p_n}} < 1 \textrm{ for large enough } n.
\end{align}
Observe that
\begin{align}
	\Rninv - \Rn'^{-1} & = \Rninv\big(\Rn' - \Rn\big)\Rn'^{-1} \\ & = \pm \Rninv E_{ij}\Rn'^{-1}/np_n.
\end{align}
This enables us to bound the difference in $f$:
\begin{align}
	& |f(\Lnone) - f(\Lnone')| \\ & = \frac{1}{n} \Big| \trace\left( \Rninv \right) - \trace\left( \Rn'^{-1} \right) \Big| \\
		& = \frac{1}{n^2p_n} \Big| \trace\left( \Rninv E_{ij} \Rn'^{-1} \right) \Big| \\
		& \leq \frac{1}{n^2p_n} \frobnorm{\Rninv (e_i - e_j)}\frobnorm{\Rn'^{-1} (e_i - e_j)} \\
		& \leq \frac{2}{n^2p_n} \opnorm{\Rninv}\opnorm{\Rn'^{-1}} \\
		& \leq \frac{2}{n^2p_n} \frac{1}{\left(1 - 5.01 \sqrt{\frac{\log n}{n p_n}}\right)^2} \leq \frac{2.01}{n^2p_n} \triangleq c.
\end{align}

We may now apply Lemma~\ref{lem:mcdiarmid}. For all $\lambda, \alpha > 0$,
\begin{align}
	\Pr\big[ \left| X_n - \expect X_n \big| \geq \lambda \right] & \leq 2\exp\left( \frac{-\lambda^2}{n(n-1)\left(\frac{2.01}{n^2 p_n} + \frac{n^2}{6}\alpha \right)^2} \right) + \frac{n(n-1)}{\alpha}\frac{3.01}{n^{11}}.
\end{align}
We have the freedom to choose $\lambda$ and $\alpha$. Let $\alpha = 1/n^5$. Then, for large $n$,
\begin{align}
	\Pr\big[ \left| X_n - \expect X_n \big| \geq \lambda \right] & \leq 2\exp\left( -(\lambda np_n/2.02)^2   \right) + \frac{3.01}{n^4}.
\end{align}
Finally, choose $\lambda > 0$ such that $-(\lambda np_n/2.02)^2 = \log \epsilon$ to conclude the proof.

\section*{Acknowledgment}

We thank Amit Singer for suggesting this collaboration, and we thank him as well as Bal\'azs Gerencs\'er and Romain Hollanders for interesting discussions. We are indebted to and thank the reviewers for identifying shortcomings in the original proofs. This paper presents research results of the Belgian Network DYSCO, funded by the Interuniversity Attraction Poles Programme initiated by the Belgian Science Policy Office. NB is an FNRS research fellow.

\bibliographystyle{elsarticle-num}
\bibliography{biblio}

\end{document}